\documentclass[10pt, aps,prc,twocolumn,superscriptaddress,preprintnumbers,
amsmath, 
floatfix,
longbibliography,
nofootinbib
]{revtex4-1}
\usepackage[T1]{fontenc}
\usepackage[utf8x]{inputenc} 
\usepackage{adjustbox}          
\usepackage[caption=false]{subfig}
\usepackage{url}
\usepackage{color}
\usepackage{float}
\usepackage[pdftex,colorlinks=true, linkcolor = blue, citecolor=blue,urlcolor=blue, bookmarksnumbered=true, bookmarksopen=true]{hyperref}
\usepackage{longtable}
\usepackage{amsfonts}
\usepackage{wrapfig,bm}
\usepackage[normalem]{ulem}
\newcommand{\beq}{\begin{equation}}
\newcommand{\eeq}{\end{equation}}
\newcommand{\bea}{\begin{eqnarray}}
\newcommand{\eea}{\end{eqnarray}}

\begin{document}

\title{Restoring Broken Symmetries for Nuclei and Reaction Fragments}
  
\author{Aurel Bulgac}%
\email{bulgac@uw.edu}%
\affiliation{Department of Physics,%
  University of Washington, Seattle, Washington 98195--1560, USA}
  
\date{\today}

\begin{abstract}

In typical microscopic approaches, particularly when pairing correlations are present, 
nuclei and nuclear fragments do not have well defined quantum numbers and 
symmetries should be restored. I present here a formalism for the simultaneous projection of total particle numbers 
of a nucleus, particle number of reaction fragments, and of the reaction fragment intrinsic spins and of their correlation, and also
for their symmetry restored densities and total energies. These formulas for the symmetry restored quantities, are free of any 
singularities, unlike those in the previously introduced  prescriptions.

\end{abstract}

\preprint{NT@UW-21-12}

\maketitle

\section{Introduction} \label{sec:I}

The problem of restoring broken symmetries within mean field treatments of nuclear systems is decades old, 
see monograph~\cite{Ring:2004} and older references therein, and new studies are published on an almost 
constant pace over the years, see many references to more recent 
studies~\cite{Anguiano:2001,Robledo:2007,Dobaczewski:2007,Lacroix:2009,Bender:2009,Duguet:2009,Bally:2021}.
Essentially all studies published so far treat
the case of either a Hartree-Fock (HF) or a Hartree-Fock-Bogoliubov (HFB)  type of generalized Slater determinant. 
Such a generalized Slater determinant is typically used to minimize the total energy of a nucleus, either before or  
after projection, within a mean field approach and from that procedure one extracts the restored symmetry 
nucleus wave functions. 

This symmetry restored wave function in either static or time-dependent formulation of the framework
is of the typical Generator Coordinate Method~\cite{Hill:1953,Griffin:1957,Goeke:1980,Verriere:2020}.
With the emergence of the  Density Functional Theory (DFT) however, the role of the (generalized) Slater determinant
was replaced by the (generalized) number densities, in which case the nuclear energy density functionals (NEDF) 
is not defined as an expectation value of a many-body Hamiltonian, but as an expectation of an 
energy density functional, which depends on several  one-body densities. 
Trying to apply the HF(B) projection techniques to DFT studies leads to a number of difficulties. 
Some of these difficulties are discussed  in 
Refs.~\cite{Anguiano:2001,Robledo:2007,Dobaczewski:2007,Lacroix:2009,Bender:2009,Duguet:2009,Bally:2021}.

The approach discussed here is based entirely on a treatment of strongly interacting many-fermion systems within the DFT framework,
 see Refs.~\cite{Bulgac:2007,Bulgac:2013a,Bulgac:2019,Bulgac:2019a} and references therein. 
The restoration of broken symmetries in case of DFT was discussed earlier~\cite{Bulgac:2019a} and it will be discussed 
in detail in this paper. The physical justification of such an approach was discussed earlier in 
Ref.~\cite{Bulgac:2010}, where a quantization of a semi-classical level was suggested, 
which can be easily converted into the projection technique discussed here.    
Unlike the approaches based on the generalized Wick theorem applied to generalized Slater
determinants and evaluation of the total energy, the present approach is free of singularities, 
see Section~\ref{sec:VII}.  A few of the results discussed here have been 
briefly discussed in Ref.~\cite{Bulgac:2019a} and a few inaccuracies in that paper are corrected here.

This is a formal paper, where I derive a series of new formulas, not discussed previously in literature,  
needed in order to restore particle and rotation broken symmetries.  
In Section~\ref{sec:II} I review some needed known facts.
In Section~\ref{sec:III} I describe how to double project the total particle number and the fragment particle number.
In Section~\ref{sec:IV} I describe how to construct particle projected number and anomalous  densities. 
In Sections~\ref{sec:V} and \ref{sec:VI} I show how to simplify the particle projection in the canonical basis. 
In Section~\ref{sec:VI} I present formulas for number and anomalous densities and for the number 
projected total energy.  
In Section~\ref{sec:VII} I describe how to simultaneously project the total particle number and the particle number 
of a reaction fragment.   In Section~\ref{sec:IX} I develop formulas for 
double projection of the total and fragment number density. In Section~\ref{sec:X} I show how to  project the
total particle and reaction fragment particle along with the intrinsic spins of the fragments and their correlations.
The particular case of total and fragment particle numbers, the intrinsic fragment spins, and the total relative 
orbital momentum are discussed in Section~\ref{sec:XI}.  The last Section~\ref{sec:XII} is devoted to the 
discussion of some numerical aspects. A number of formulas discussed here have been recently used in 
Refs.~\cite{Bulgac:2021,Bulgac:2021b}.

These formulas presented here were developed for fission applications, but they can be used for heavy-ion reactions as well, 
with some small adjustments. The presentation here is restricted to systems with even particle parity, but its extension
appears to be simple.

\section{\bf Structure of a generalized Slater determinant} \label{sec:II}

The creation and annihilation quasi-particle operators are represented as~\cite{Ring:2004}
\begin{align}
\!\!\!\!\!\! &\alpha_k^\dagger  = 
\int d\xi\left [ {\textrm u}_k(\xi) \psi^\dagger (\xi) + {\textrm v}_k(\xi) \psi (\xi)\right ], \label{eq:a0}\\
\!\!\!\!\!\! &\alpha_k= 
\int d\xi\left [ {\textrm v}_k^*(\xi) \psi^\dagger (\xi ) + {\textrm u}_k^*(\xi) \psi (\xi)\right ], \label{eq:b0}
\end{align}
and the reverse relations
\begin{align} 
&\psi^\dagger (\xi) = \sum_k \left [ {\textrm u}^*_k(\xi)  \alpha^\dagger _k  
                                            + {\textrm v}_k(\xi)\alpha_k \right ], \label{eq:p1}\\
&\psi(\xi) =                \sum_k \left [ {\textrm v}^*_k(\xi)\alpha^\dagger_k
                                            + {\textrm u}_k(\xi)\alpha_k \right ], \label{eq:p2}
\end{align}
where $\psi^\dagger (\xi )$ and $ \psi (\xi)$ are the field operators 
for the creation and annihilation of a particle with coordinate $\xi$.
The normal number (Hermitian $n=n^\dagger$ ) and anomalous (skew symmetric $\kappa =-\kappa^T$) densities are
\begin{align}
n(\xi,\xi') &= \langle \Phi|\psi^\dagger(\xi')\psi(\xi)|\Phi\rangle \label{eq:number} \\
&= \sum_k {\textrm v}_k^*(\xi) {\textrm v}_k(\xi')=\sum_{l=n,\bar{n}} v_l^2  \phi_l^*(\xi) \phi_l(\xi'),\nonumber \\
\kappa(\xi,\xi') &= \langle \Phi|\psi(\xi')\psi(\xi)|\Phi\rangle \label{eq:numbera} \\
&=\sum_k {\textrm v}_k^*(\xi){\textrm u}_k(\xi') = \sum_{l=n,\bar{n}} u_lv_l \phi_l^*(\xi) \phi_{\bar{l}}^*(\xi'), \nonumber \\
&  \int d\xi \phi_k^*(\xi) \phi_l(\xi)=\delta_{kl},
\end{align}
with $u_l^2+v_l^2=1$, $0\le u_l=u_{\bar{l}}\le 1$,  $0\le v_l=-v_{\bar{l}}\le 1$, and $n$ and $\bar{n}$ denote time-reversed states 
in the canonical representation~\cite{Bloch:1962,Ring:2004,Balian:1969}, and where
\begin{align}
& \alpha_k|\Phi\rangle =0, \quad |\Phi\rangle = {\cal N} \prod_k\alpha_k|0\rangle , \quad \langle \Phi|\alpha_k\alpha_l^\dagger|\Phi\rangle =\delta_{kl},
\end{align}
where ${\cal N}$ is a normalization factor, determined up to an arbitrary phase, and assuming that $\alpha_k|0\rangle \neq 0$ for any $k$. 
In case any $\int d\xi |{\textrm v}_k(\xi)|^2 = 0$ or $\alpha_k|0\rangle =0$ the corresponding factor ${\alpha}_k$  
is skipped. Here the discussion will be explicitly limited to systems with an even particle number parity, 
as the extension to the general case is trivial~\cite{Ring:2004}.

Here I will elaborate at first on details of the projection technique developed in Ref.~\cite{Bulgac:2019a}, which were not discussed before.
The particle projection on a fragment of the system is performed with the help of the unitary operator, introduced earlier in Ref.~\cite{Simenel:2010} 
\begin{align}
&\hat{P}^\Theta(\eta) = e^{i\eta\int d\xi \Theta(\xi) \psi^\dagger(\xi)\psi(\xi)}=e^{i\eta\hat{N}^\Theta}, \\
&\hat{N}^\Theta =\int d\xi\Psi^\dagger(\xi)\psi(\xi)\Theta(\xi), \\
& \Theta^2(\xi)=\Theta(\xi),\quad   \hat{P}^\Theta(\eta)\hat{P}^\Theta(-\eta)=1, \quad \eta\in[-\pi,\pi]
\end{align}
$\Theta(\xi)$ is the Heaviside function
and for all non-negative integer particle numbers
\begin{align}
|\Phi^\Theta(N)\rangle = \int_{-\pi}^\pi \frac{d\eta}{2\pi}e^{-i\eta N} \hat{P}^\Theta(\eta)|\Phi\rangle .
\end{align}
is the component of the wave function $|\Phi\rangle$ with exactly $N$ particles in the space region where $\Theta(\xi) =1$. 

One can easily show that under the transformation with this operator the field and quasiparticle operators change according to the rules
\begin{align}
&\psi^\dagger(\xi,\eta) = \hat{P}^\Theta(\eta)\psi^\dagger(\xi) \hat{P}^\Theta(-\eta)=e^{i\eta\Theta(\xi)}\psi^\dagger(\xi), \label{eq:eta5}\\
\!\!\!\!\!\!\!\!\!\!\!\! &\tilde{\alpha}_k (\eta)=
\int \!\!\!d\xi\left [ e^{i\eta\Theta(\xi)}{\textrm v}_k^*(\xi) \psi^\dagger (\xi ) + e^{-i\eta\Theta(\xi)}{\textrm u}_k^*(\xi) \psi (\xi)\right ].\nonumber
\end{align}
It is easy to show that
\begin{align}
\{\tilde{\alpha}_k^\dagger(\eta),\tilde{\alpha}_l(\eta)\}=\delta_{kl},\quad \{\tilde{\alpha}_k(\eta),\tilde{\alpha}_l(\eta)\}=0.
\end{align}
This implies that when $\Theta(\xi)\equiv 1$ the components of the quasiparticle wave functions (qpwfs) change as
\begin{align}
[ {\textrm v}_k^*(\xi), {\textrm u}_k^*(\xi)]
 \rightarrow  
 [e^{i\eta\Theta(\xi)}{\textrm v}_k^*(\xi), e^{-i\eta\Theta(\xi)}{\textrm u}_k^*(\xi)]. \label{eq:trans} 
\end{align} 
and correspondingly the new vacuum is (assuming that for all   $\tilde{\alpha}_k|0\rangle >0$) 
\begin{align}
 |\tilde{\Phi}(\eta)\rangle ={\cal N}\prod_k\tilde{\alpha}_k(\eta) |0\rangle =\hat{P}^\Theta(\eta)|\Phi\rangle  . \label{eq:Phis}
 \end{align}
   
 In the case $2\Omega = 4$ the wave function $|\Phi\rangle$ will have 4-particle, 2-particle,  and 0-particle components. 
 A typical 2-particle component  arising from 
\begin{align}
\int d\xi_1 d\xi_2 d\xi_3 d\xi_4 &\; 
{\textrm u}^{*}_{1}(\xi_1){\textrm v}^{*}_{2}(\xi_2)  {\textrm v}^{*}_{3}(\xi_3){\textrm v}^{*}_{4}(\xi_4) \nonumber \\
& \psi(\xi_1)\psi^\dagger(\xi_2)\psi^\dagger(\xi_3)\psi^\dagger(\xi_4)|0\rangle \nonumber 
\end{align} 
has the structure
\begin{align}
 & = \int d\xi  {\textrm u}^{*}_{1}(\xi){\textrm v}^{*}_{2}(\xi) \int d\xi_1 d\xi_2 
 {\textrm v}^{*}_{3}(\xi_1){\textrm v}^{*}_{4}(\xi_2)   \psi^\dagger(\xi_1) \psi^\dagger(\xi_2)|0\rangle \nonumber\\
&     - \int d\xi  {\textrm u}^{*}_{1}(\xi){\textrm v}^{*}_{3}(\xi)  \int d\xi_1 d\xi_2 
{\textrm v}^{*}_{2}(\xi_1){\textrm v}^{*}_{4}(\xi_2)   \psi^\dagger(\xi_1) \psi^\dagger(\xi_2)|0\rangle \nonumber \\ 
 &   + \int d\xi  {\textrm u}^{*}_{1}(\xi){\textrm v}^{*}_{4}(\xi) \int d\xi_1 d\xi_2 
 {\textrm v}^{*}_{1}(\xi_1){\textrm v}^{*}_{4}(\xi_2)   \psi^\dagger(\xi_1) \psi^\dagger(\xi_2)|0\rangle \nonumber .
 \end{align}  
There are two more contributions to the 2-particle component arising from the terms containing the combinations of field operators 
 $ \psi^\dagger(\xi_1)\psi(\xi_2)\psi^\dagger(\xi_3)\psi^\dagger(\xi_4)$ and $ \psi^\dagger(\xi_1)\psi^\dagger(\xi_2)\psi(\xi_3)\psi^\dagger(\xi_4)$. 

  After applying the operator $\hat{P}^\Theta(\eta)$ on the above 2-particle component only the quasiparticle ${\textrm v}$-components change as
 ${\textrm v}_k^*(\xi) \rightarrow e^{i\eta\Theta(\xi)}{\textrm v}_k^*(\xi)$, but only  for terms with factors like $\int d\xi {\textrm v}_k(\xi)\psi^\dagger(\xi)$.  
 Terms containing factors of the type $\int d\xi {\textrm u}_k(\xi)\psi(\xi)$ do not survive after normal ordering. The 
 terms like $ \int d\xi  {\textrm u}^{*}_{k}(\xi){\textrm v}^{*}_{l}(\xi) $ are left invariant by either by transformation 
 Eq.~\eqref{eq:trans} or by the operator $\hat{P}^\Theta(\eta)$. 
  
 According to the analysis performed above on the example of $2\Omega =4$ only the overlaps between the ${\textrm v}$-components
of the qpwfs  in the Onishi-Yoshida~\cite{Onishi:1966,Ring:2004} formula are changed, namely
\begin{align}
\langle \Phi|\hat{P}^\Theta(\eta)|\Phi\rangle &= 
\sqrt{\det{[ \langle {\textrm u}_k|{\textrm u}_l\rangle + \langle {\textrm v}_k|e^{i\eta\Theta}|{\textrm v}_l\rangle]}}\nonumber\\
&= \sqrt{\det{[ \delta_{kl} + (e^{i\eta}-1)\langle {\textrm v}_k|\Theta|{\textrm v}_l\rangle]}}.
\end{align}
One should note, that no overlaps of the type $ \int d\xi  {\textrm u}^{*}_{k}(\xi){\textrm v}^{*}_{l}(\xi)\Theta(\xi) $ 
appear in the Onishi-Yoshida overlap formula, which otherwise might have led to spurious terms.

\section{Double projection of fragment particle number and also overall particle number} \label{sec:III}

When evaluating the particle number of a fragment one should remember that its particle number distribution is affected by the 
uncertainty in the particle number in the total many-body wave function. Let me consider the projection of the total particle number
\begin{align}
&e^{i\eta_0\hat{N}}|\Phi\rangle =\sum_{n=0}^{\Omega}a_{2n}e^{2in\eta_0}|\Phi_{2n}\rangle, \label{eq:ff0}\\
&\sum_{n=0}^\Omega |a_{2n}|^2 =1, \quad \hat{N}=\int d\xi\psi^\dagger(\xi)\psi(\xi),
\end{align}
where $n=N$ are non-negative integers and $\Phi_{2n}$ are linear combinations of ordinary Slater 
determinants for exactly $N=2n$ particles. Since only even $2n\eta_0$ frequencies are present one can limit 
the integral over the interval $\eta_0\in[-\pi/2,\pi/2]$.

The wave function~\eqref{eq:Phis} constructed for $\Theta\equiv 1$
\begin{align}
&|\tilde{\Phi}(\eta_0)\rangle =\hat{P}^\Theta(\eta_0)|\Phi\rangle={\cal N}\prod _k\tilde{\alpha}_k(\eta_0)|0\rangle,\label{eq:ff}
\end{align}
 with the operators
 \begin{align}
&\tilde{\alpha}_k(\eta_0)=
\int d\xi\left  [ e^{i\eta_0}{\textrm v}_k^*(\xi) \psi^\dagger (\xi ) + e^{-i\eta_0}{\textrm u}_k^*(\xi) \psi (\xi)\right ].
\end{align}
has according to Onishi-Yoshida formula the overlap
\begin{align}
&\langle\Phi|\tilde{\Phi}(\eta_0)\rangle = \sqrt{\det{ [e^{-i\eta_0} \langle {\textrm u}_k|{\textrm u}_l\rangle
+e^{i\eta_0}\langle{\textrm v}_k|{\textrm v}_l\rangle] }}, \\
&=e^{-i\eta_0\Omega} \sqrt{\det{ [\delta_{kl} + (e^{2i\eta_0}-1) \langle{\textrm v}_k|{\textrm v}_l\rangle]}}
\end{align} 
with both positive and negative frequencies $e^{in\eta_0 }$, 
\begin{align}
\langle\Phi|\tilde{\Phi}(\eta_0)\rangle= e^{-i\eta_0\Omega} \sum_{m=0}^{2\Omega} \tilde{a}_{2m} e^{2im\eta_0}. \label{eq:N0}
\end{align}
From the arguments presented in Sections~\ref{sec:V} and \ref{sec:VI} and from 
our numerical simulations~\cite{PPFF:2021} as well it follows that the 
frequency spectrum lies predominantly in the interval $[-\Omega,\Omega]\eta_0$,
unlike the natural expansion 
Eq.~\eqref{eq:ff0}, where only the expected terms with $0\le N=2n\le 2\Omega$ are present. 
In the particular case of ordinary Slater determinant with exactly $N$-particles one obtains using Onishi-Yoshida formula
\begin{align}
\langle \Phi|\tilde{\Phi}(\eta_0)\rangle = e^{-i\eta_0\Omega}e^{i\eta_0N},
\end{align}
since $\langle{\textrm u}_k |{\textrm u}_k\rangle +  \langle{\textrm v}_k|{\textrm v}_k\rangle=1$  
and there are exactly $N$ overlaps $\langle{\textrm v}_k|{\textrm v}_k\rangle =1$, while the rest $2\Omega-N$ such overlaps vanish.
Thus, using Onishi-Yoshida overlap formula results in an incorrect frequency spectrum, a situation which can 
be quite easily rectified as suggested below.

It is useful to introduce a different set of annihilation operators~\cite{Bulgac:2019a}
\begin{align}
&\alpha_k(\eta_0)=\int d\xi [ e^{2i\eta_0} {\textrm v}_k^*(\xi)\psi^\dagger(\xi)+ {\textrm u}_k(\xi)\psi(\xi)] \\
=&e^{-i\eta_0}\tilde{\alpha}_k(\eta_0)=\sum_l [A_{kl}(\eta_0) \alpha_l +B_{kl}(\eta_0)\alpha^\dagger_l], \\
&A_{kl}(\eta_0)  = \delta_{kl} + (e^{2i\eta_0}-1) \int d\xi  {\textrm v}_k^*(\xi){\textrm v}_l(\xi)\,  \\
&B_{kl}(\eta_0) =  (e^{2i\eta_0}-1) \int d\xi {\textrm v}^*_k(\xi){\textrm u}_l^*(\xi), 
\end{align}
with the new associated qpwfs
\begin{align} 
 [{\textrm v}_k^*(\xi), {\textrm u}_k^*(\xi)] \rightarrow
 [e^{i2\eta_0}{\textrm v}_k^*(\xi),{\textrm u}_k^*(\xi)]\label{eq:muv1}
 \end{align} 
and
\begin{align}
&|\Phi(\eta_0)\rangle = {\cal N}\prod_k\alpha_k(\eta_0)|0\rangle.
\end{align}
On can then easily see that
\begin{align}
\langle \Phi | \Phi(\eta_0)\rangle = e^{ i\eta_0\Omega } \langle\Phi|\tilde{\Phi}(\eta_0)\rangle 
= \sum_{n=0}^\Omega  a_{2n} e^{2in\eta_0},
\end{align}
similarly to  Eq.~\eqref{eq:ff0} and also that
\begin{align}
\langle \Phi|\Phi(\eta_0)\rangle = e^{i\eta_0N}
\end{align}
for the case of an ordinary Slater determinant for $N$-particles one obtains the correct result. 
These conclusions are also confirmed in Sections~\ref{sec:V} and \ref{sec:VI},
where an analysis is performed using the canonical basis.
Numerical simulations also show that $\max_N |a_{N}|^2$ occurs, as naturally expected, 
for $N\approx \langle \Phi|\hat{N}|\Phi\rangle $, see also Ref.~\cite{Bulgac:2019a}. 

It then follows that the projected overlap on the total particle number $N$ wave function
\begin{align}
&\langle \Phi |\Phi_N(\eta^\text{F})\rangle = \int_{-\pi}^{\pi}  \frac{d\eta_0}{2\pi} e^{ -i \eta_0 N } \langle \Phi |\Phi(\eta_0,\eta^\text{F})\rangle,\label{eq:nnF1} \\
&\langle\Phi |\Phi(\eta_0,\eta^\text{F})\rangle = {\cal N}(\eta_0,\eta^\text{F})\label{eq:nnF2}\\
&\times\langle\Phi| \prod_k \int d\xi\left [ 
e^{ 2i\eta_0}e^{i\eta^\text{F}\Theta^\text{F}(\xi) }  {\textrm v}_k^*(\xi) \psi^\dagger (\xi ) + {\textrm u}_k^*(\xi) \psi (\xi)\right ]|0\rangle \nonumber
\end{align}
is a sum of overlaps of  (ordinary) Slater determinants for exactly $N$-particles, where $0\le N\le 2\Omega$ is even. 

As I discussed in the previous section,
in $|\Phi\rangle = {\cal N}\prod_{k=1}^{2\Omega}\alpha_k|0\rangle$ only terms with an even number 
creation operators $\psi^\dagger(\xi)$ and no annihilation operators $\psi(\xi)$ survive after normal ordering. The integration over the angle $\eta_0$ 
selects only terms with exactly $N$ creation operators $\psi^\dagger(\xi)$ from $|\Phi(\eta_0,\eta^\text{F})\rangle$.
In order to correctly evaluate the particle number in a reaction 
fragment one has to perform a double particle number projection, on the total particle number $N$ and on the fragment particle 
(integer) number $N^\text{F}$, where $0\le N^\text{F}\le N$. 

In  order to accurately determine the particle number in a fission fragment (FF) one has to perform a double particle 
projection~\cite{Scamps:2013,Verriere:2021,Verriere:2021a}, the first projection to fix the total particle number in the 
fissioning nucleus and the second projection to determine the particle number in the FF. One has thus to consider the overlap
\begin{align}
&\langle \Psi |\Psi (\eta_0, \eta^\text{F})\rangle 
=\sqrt{\det{[ \delta_{kl} + \langle \textrm{v}_k | e^{2i\eta_0} e^{i\eta^\textrm{F}\Theta^\textrm{F}} -1 |\textrm{v}_l\rangle] } } \nonumber \\
& \!\!\!\!\!\!\!= \sqrt{ \det{ [\delta_{kl} + (e^{2i\eta_0}-1)O_{kl} +e^{2i\eta_0}(e^{i\eta^\textrm{F}}-1) O_{kl}^\textrm{F} ]  }} , \label{eq:NNf}\\
&O_{kl}= 
\langle {\textrm v}_k |  {\textrm v_l}\rangle,\quad 
O_{kl}^\textrm{F}=  \langle {\textrm v}_k | \Theta^\text{L,H} | {\textrm v_l}\rangle,\\
& O_{kl}^\text{H}+O_{kl}^\text{L}= O_{kl}, \quad \text{if} \quad \Theta^\text{L}+\Theta^\text{H} =1,
\end{align}
and where $\Theta^\text{F}=\Theta^\text{L,H}$ selects the spatial region of either the heavy (H) or of the light (L) FF.
The double particle projection is required as the initial state does not have a well defined particle number. 
Since $N=N^\text{L}+N^\text{H}$ the probability distributions for the two FFs are related $P(N,N^\text{L})=P(N,N-N^\text{H})$, where
\begin{align}
P(N,N^\text{F}) = & \int_{-\pi/2}^{\pi/2}  \frac{ d\eta_0 }{ \pi }\int_{-\pi}^\pi\frac{ d\eta^\text{F} }{ 2\pi } \\
&\times \textrm{Re}[\langle \Psi |\Psi (\eta_0, \eta^\text{F})\rangle e^{-i\eta_0N-i\eta^\text{F}N^\text{F}}]\nonumber
\end{align} 
The particle probability distribution in a fragment is given
by the conditional probability 
\begin{align}
P_N( N^\text{F} ) = \frac{ P(N,N^\text{F} ) }{ \sum_{N^\text{F}=0}^N  P(N,N^\text{F})}.
\end{align} 

In case of a reaction between two superfluid nuclei one needs to perform a triple projection, on both initial partners 
and one on the final fragment.

The attentive reader has noticed that in Ref.~\cite{Bulgac:2019a} it was argued that for a FF particle projection, 
where the projection on the total particle number was not considered, one should use the overlap
\begin{align}
&\langle \Psi |\Psi ( \eta^\text{F})\rangle 
=\sqrt{\det{[ \delta_{kl} + \langle \textrm{v}_k |e^{i\eta^\textrm{F}\Theta^\textrm{F}} -1 |\textrm{v}_l\rangle] } } \nonumber \\
& \!\!\!\!\!\!\!= \sqrt{ \det{ [\delta_{kl} +(e^{i\eta^\textrm{F}}-1) O_{kl}^\textrm{F} ]  }}. 
\end{align}
Since the projection on the total particle number selects in Eq.~\eqref{eq:NNf} overlaps of ordinary Slater determinants, 
the projection of the FF particle number can proceed following the procedure outlined above, see Eqs.~\eqref{eq:nnF1} and (\ref{eq:nnF2}),
as it was established earlier in the literature~\cite{Simenel:2010,Bulgac:2019a}.

\section{ Projecting the particle number for an arbitrary one-body observable}\label{sec:IV}

Here I will derive a formula for a particle average of the operator 
$\hat{Q} = \int d\xi d\xi' \langle \xi |Q|\xi' \rangle \psi^\dagger (\xi) \psi(\xi')$.
Consider at first the transformation
\begin{align}
&{\textrm u}_k(\xi,\epsilon) = {\textrm u}_k(\xi),\quad  {\textrm v}_k(\xi,\epsilon) = e^{2\epsilon \hat{Q}}{\textrm v}_k(\xi), \label{eq:uvq}
\end{align}
one can show that
\begin{align}
&\left .   \frac{d  \langle \Phi | \Phi(\epsilon)\rangle }{d \epsilon } \right |_{\epsilon =0} 
=\lim_{\epsilon\rightarrow 0}\frac{ \sqrt{ \det{[ \delta_{kl} + \langle {\textrm v}_k| e^{2\epsilon \hat{Q} }-1|{\textrm v}_l\rangle]  }   } }{\epsilon}\nonumber \\
& =  \sum_k \langle {\textrm v}_k|\hat{Q}|{\textrm v}_k\rangle =  \langle \Phi | \hat{Q} |\Phi \rangle,
\end{align} 
where $|\Phi(\epsilon)\rangle $ was constructed  with qpwfs~\eqref{eq:uvq}. 
Since one needs the ``deformed'' quasi-particle wave functions with an accuracy ${\cal O}(\epsilon)$ only one can use
$1+2\epsilon {\hat Q}$ instead of $e^{2\epsilon \hat{Q}}$. In this case the transformation of the quasi-particle wave functions is
\begin{align}
&{\textrm u}_n(\xi,\epsilon) = {\textrm u}_n(\xi),\\
&{\textrm v}_n(\xi,\epsilon) = \int\!\!d\xi' [ \delta(\xi-\xi')+2\epsilon \langle \xi |Q|\xi' \rangle ]{\textrm v}_n(\xi').\nonumber
\end{align}
The number density matrix - and in a similar manner the anomalous density, see below -  is naturally defined as a functional derivative, 
see \textcite{Negele:1988,Furnstahl:2005},
\begin{align}
n(\xi,\xi') = \frac{\delta q}{\delta \langle \xi |Q|\xi' \rangle}, \, \text{where}\, q = \langle \Phi | \hat{Q} |\Phi \rangle
\end{align}
This definition of the number density matrix, as the functional 
derivative of the partition function with respect to an arbitrary external field 
and which is widely used in quantum field theory for decades,  is the main difference between the broken symmetry 
restoration framework described here and those introduced in previous approaches. The density matrix is thus naturally defined as the 
response or the measurement due to an appropriately chosen weak external probe acting on the system. 

The normal particle projected one-body density can be calculated as the variational derivative
\begin{align}
n(\xi,\xi'|\eta_0) = \frac{\delta q(\eta_0)}{\delta \langle \xi |Q|\xi' \rangle}
\end{align}
where, in order to evaluate $q(\eta_0)$ one should use now the overlap 
\begin{align}
\langle \Phi | \Phi(\epsilon,\eta_0)\rangle=  \sqrt{ \det{[ \delta_{kl} + \langle {\textrm v}_k| e^{2i\eta_0}e^{2\epsilon \hat{Q} }-1|{\textrm v}_l\rangle]  }   } 
\end{align}
and thus
\begin{align}
&n(\xi,\xi'|\eta_0)          = \langle\Phi|\Phi(\eta_0)\rangle e^{2i\eta_0} \sum_{kl}  {\textrm v}_k^*(\xi){\textrm v}_l(\xi')\, a_{lk}(\eta_0). \label{eq:nm1}
\end{align}  
The matrix $a_{kl}(\eta_0)$ is the inverse of the matrix $A_{kl}(\eta_0)$
\begin{align}
&A_{kl}(\eta_0) = [\delta_{kl} +(e^{2i\eta_0}-1)  \langle {\textrm v}_k | {\textrm v}_l\rangle], \\
&\sum_l A_{kl}(\eta_0)a_{lm}(\eta_0) = \delta_{km}.
\end{align}

In the case of the anomalous density $\kappa(\xi,\xi'|\eta_0)$ one would have to consider a transformation different from
Eq.~\eqref{eq:uvqe}, namely the transformation
\begin{align} 
&u_n(\xi,\epsilon,\eta_0) = u_n(\xi) +   2\epsilon \int \!\! d\xi' \langle \xi|\Delta |\xi'\rangle  e^{2i\eta_0}v_n(\xi'),\nonumber\\
&v_n(\xi,\epsilon,\eta_0) = v_n (\xi)  
\end{align}
in order to construct $|\Phi(\epsilon,\eta_0)\rangle$ and follow the same steps as in the case of a normal operator 
$\hat{Q}$ outlined above and obtain for the anomalous density Eq.~\eqref{eq:numbera}
 \begin{align}
&\kappa(\xi,\xi'|\eta_0) = \langle\Phi|\Phi(\eta_0)\rangle e^{2i\eta_0} \sum_{lk}  {\textrm v}_k^*(\xi){\textrm u}_l(\xi')\, a_{lk}(\eta_0) \label{eq:nm2}.
\end{align}  
These formulas simplify significantly in the canonical basis, see Section~\ref{sec:VI}.

\section{Canonical basis}\label{sec:V}

The calculation of the particle projected averages are greatly simplified in the canonical basis.
After diagonalizing the overlap $O_{kl}=\langle {\textrm v}_k|{\textrm v}_l\rangle$ of the ${\textrm v}$-components 
the new qpwfs satisfy the relations
\begin{align} 
\langle\tilde{{\textrm v}}_k|\tilde{{\textrm v}}_l\rangle = n_k\delta_{kl} \label{eq:noc}
\end{align}
it follows that the overlap matrix of the ${\textrm u}_k$-components is also diagonal
\begin{align} 
\langle\tilde{{\textrm u}}_k|\tilde{{\textrm u}}_l\rangle = (1-n_k)\delta_{kl},
\end{align}
and the average particle number is given by 
\begin{align} 
N = \sum_k n_k.
\end{align}
The occupation probabilities $n_k=\langle\tilde{{\textrm v}}_k|\tilde{{\textrm v}}_k\rangle$ are different from 
$\langle{\textrm v}_k|{\textrm v}_k\rangle$, even though their sums add to the same total particle number $N$, 
due to invariance of the trace of a matrix.
The number of ${\textrm v}_k$-components is $2\Omega =2N_xN_yN_z$ for neutrons and protons respectively. 
In an infinite box $2\Omega = \infty$. 

It is useful to introduce the unitary transformation, and correspondingly the set of eigenvectors, which diagonalizes $O_{kl}$
\begin{align}
&\sum_lO_{kl}{\cal U}_{lm} = {\cal U}_{km}n_m, \quad \sum_n{\cal U}^*_{km}{\cal U}_{kn} = \delta_{mn},\\
&{\textrm v}_k(\xi)=\sum_m{\cal U}_{km}\tilde{{\textrm v}}_m(\xi), 
\quad \tilde{\textrm v}_n(\xi)=\sum_l{\cal U}_{ln}^*{\textrm v}_l(\xi)\\
&O_{kl} = \sum_m{\cal U}_{km}n_m{\cal U}^*_{lm}.
\end{align}

In the canonical basis the overlap for the double particle projection Eq.~\eqref{eq:NNf} acquire the simpler form
\begin{align}
&\langle \Psi |\Psi (\eta_0, \eta^\text{F})\rangle \label{eq:NNf2}\\
& \!\!\!\!\!\!\!= \sqrt{ \det{\left [  [1+ (e^{2i\eta_0}-1)n_k]\delta_{kl} +e^{2i\eta_0}(e^{i\eta^\textrm{F}}-1) \tilde{O}_{kl}^\textrm{F}\right ]  }} , \nonumber \\
& \tilde{O}_{kl}^\textrm{F}=  \langle \tilde{\textrm v}_k | \Theta^\text{L,H} | \tilde{\textrm v_l}\rangle.
\end{align}

The overlap $\langle\tilde{{\textrm v}}_k(t)|\tilde{{\textrm v}}_l(t)\rangle $ does not remain diagonal as a function 
of time in a time-dependent evolution. For that reason the simplified formulas for the number projected quantities should be 
derived in the canonical basis determined at the time when the corresponding observables are needed.

\section{Textbook definition of the canonical basis} \label{sec:VI}

Since the eigenvalues of the matrix ${\cal O}_{kl}$ are double degenerate and can always choose the canonical qpwfs 
$\tilde{\textrm u}_k(\xi), \tilde{\textrm v}_k(\xi)$ of the textbook form~\cite{Ring:2004}
\begin{align}
|\Phi\rangle =  {\cal N}\prod_{n=1}^\Omega\alpha_n\alpha_{\overline{n}} |0\rangle 
= \prod_{n=1}^\Omega(u_n+v_na^\dagger_na^\dagger_{\overline{n}})|0\rangle,
\end{align}
where
\begin{align}
 &\alpha_n=u_na_{n} - v_na^\dagger_{\overline{n}},  \quad \alpha_{\overline{n}} = u_na_{\overline{n}} +v_n a^\dagger_n,\\ 
 &a^\dagger_n=\int d\xi \phi_n(\xi)\psi^\dagger(\xi), \quad a^\dagger_{\overline{n}}=\int d\xi\phi_{\overline{n}}(\xi)\psi^\dagger(\xi),\\ 
 &\langle\phi_n|\phi_n\rangle = \langle\phi_{\overline{n}}|\phi_{\overline{n}} \rangle =1,  \quad \langle\phi_{\overline{n}}|\phi_n \rangle =0
\end{align}
and real $u_n\ge 0, \,v_n\ge 0$.

After normal ordering one obtains
 \begin{align}
 &\alpha_n\alpha_{\overline{n}} = v_n^2 a^\dagger_na^\dagger_{\overline{n}} +u_nv_n\\
                                                 &+u^2_n a_na_{\overline{n}} -u_nv_n(a^\dagger_na_n +a^\dagger_{\overline{n}}a_{\overline{n}}). \nonumber\\ 
&\frac{1}{v_n}\alpha_n\alpha_{\overline{n}}|0\rangle  = (u_n+v_n a^\dagger_na^\dagger_{\overline{n}})|0\rangle, \quad u_n^2+v_n^2=1.\label{eq:ab}
\end{align}

 After a gauge transformation $\hat{P}^\Theta(\eta)\, \alpha_n\alpha_{\overline{n}}|0\rangle$ only the 
 creation operators $a^\dagger_na^\dagger_{\overline{n}}$  in Eq.~\eqref{eq:ab} are affected by the action of $\hat{P}^\Theta(\eta)$.     
 Then the overlap 
 \begin{align} 
 & \frac{1}{v_mv_n}\langle 0|   \alpha^\dagger_{\overline{m}}\alpha^\dagger_m\, \hat{P}^\Theta(\eta)\, \alpha_n\alpha_{\overline{n}}|0\rangle\nonumber\\
 &= u_mu_n +v_mv_n \langle 0| a_{\overline{m}}a_m\hat{P}^\Theta(\eta)a^\dagger_na^\dagger_{\overline{n}}|0\rangle.
 \end{align} 
 The matrix element can be simplified 
 \begin{align}
&\langle 0| a_{\overline{m}}a_m\hat{P}^\Theta(\eta)a^\dagger_na^\dagger_{\overline{n}}|0\rangle =\\
& =\{ \,\, [\delta_{mn}+(e^{i\eta}-1)\langle \phi_m |\Theta | \phi_n \rangle ]\nonumber \\
 &\quad\times [\delta_{mn}+ (e^{i\eta}-1)\langle \phi_{\overline{m}} | \Theta |  \phi_{\overline{n}}\rangle ]\nonumber \\
 &                         -(e^{i\eta}-1)^2 \langle \phi_m |\Theta  |\phi_{\overline{n}} \rangle \langle \phi_{\overline{m}} | \Theta | \phi_n \rangle \}. \nonumber
 \end{align}
 If $\Theta(\xi)\equiv 1$ this formula simplifies
 \begin{align}
&\frac{1}{v_mv_n}\langle 0| \alpha^\dagger_{\overline{m}}\alpha^\dagger_m\hat{P}^\Theta(\eta)\alpha_n\alpha_{\overline{n}}|0\rangle = 
\delta_{mn}[u_n^2+e^{2i\eta}v_n^2]. 
\end{align}

I will introduce now the gauge transformed operators and total wave function and using Eq.~\eqref{eq:ab} one obtains
\begin{align}
&\alpha_n(\eta_0) = u_na_n-v_ne^{2i\eta_0}a^\dagger_{\overline{n}},\\
&\alpha_{\overline{n}}(\eta_0) = u_na_{\overline{n}}+v_ne^{2i\eta_0}a^\dagger_n,\\
&\alpha_n|\Phi(\eta_0)\rangle =0, \, \alpha_{\overline{n}}|\Phi(\eta_0\rangle=0,\\
&|\Phi(\eta_0)\rangle= \sum_{n=0}^\Omega a_{2n} e^{2i\eta_0n} |\Phi_{2n}\rangle,\\  
&\sum_{n=0}^\Omega |a_{2n}|^2 = 1,
\end{align}
and where $n,{\overline{n}}=1,\dots,\Omega$ and where $|\Phi_{2n}\rangle $
 are sums of (ordinary) Slater determinants for exactly $N=2n$ particles.
 The particle probability distribution is thus given by
 \begin{align}
 &P(N)=|a_{N}|^2 = 2\,{\textrm Re} \int_{0}^{\pi/2}\!\!\frac{d\eta_0}{\pi} e^{-i\eta_0N}\langle \Phi|\Phi(\eta_0)\rangle,    \\
 &\langle \Phi|\Phi(\eta_0)\rangle= \prod_{n=1}^\Omega [u_n^2+v_n^2e^{2i\eta_0}],
 \end{align} 
where the integration integral was halved, since $\langle\Phi|\Phi(\eta_0+\pi)\rangle =  \langle\Phi|\Phi(\eta_0)$.

In the case of double particle projection one introduces the quasiparticle operators
\begin{align}
&\alpha_n(\eta_0,\eta^\text{F})                  \nonumber \\
& = -v_ne^{2i\eta_0}\!\!\!\int \!\!d\xi \phi_n(\xi)                   e^{i\eta^\text{F}\Theta^\text{F}(\xi)}\psi^\dagger(\xi) +u_na_n,\\
&\alpha_{\overline{n}}(\eta_0,\eta^\text{F}) \nonumber \\
&=  v_ne^{2i\eta_0}\int \!\!d\xi \phi_{\overline{n}}(\xi)e^{i\eta^\text{F}\Theta^\text{F}(\xi)}\psi^\dagger(\xi)+u_na_{\overline{n}},
\end{align}
and the corresponding overlap has the structure
\begin{align}
&\langle \Phi|\Phi(\eta_0,\eta^{\text{F}}) \rangle \nonumber \\
=& \sqrt{\det { [\delta_{kl} + 
{\textrm v}_k{\textrm v}_l \langle \phi_k | e^{2i\eta_0} e^{i\eta^\text{F}\Theta^\text{F}}-1 | \phi_l \rangle ] }}, 
\end{align}
where $k,l$ run over both sets of $n,\overline{n}=1,\ldots ,\Omega$ and $n_{k,l}={\textrm v}^2_{k,l}$ are occupation probabilities, see Eq.~\eqref{eq:noc}. 

\section{Particle number projected  densities and total energy}  \label{sec:VII}

For any FF observables expression of the projected densities are useful. 
The densities $n(\xi,\xi'|\eta_0)$, Eq.~\eqref{eq:nm1} and  $\kappa(\xi,\xi'|\eta_0)$, Eq.~\eqref{eq:nm2} 
acquire in the canonical basis a simple form
\begin{align} 
\!\!\!\!\!\!n(\xi,\xi'|\eta_0)&=\langle \Phi|\Phi(\eta_0)\rangle  
        \sum_k \frac{\tilde{{\textrm v}}_k^*(\xi)\tilde{{\textrm v}}_k(\xi')e^{2i\eta_0}}{1+(e^{2i\eta_0}-1)n_k},\label{eq:NNF}\\
\!\!\!\!\!\!\kappa(\xi,\xi'|\eta_0)&=\langle \Phi|\Phi(\eta_0)\rangle  
        \sum_k \frac{\tilde{{\textrm v}}_k^*(\xi)\tilde{{\textrm u}}_k(\xi')e^{2i\eta_0}}{1+(e^{2i\eta_0}-1)n_k}, \label{eq:NNF1}
\end{align}
where the sum and products run over all quasiparticle states. 
The use of Eqs.~\eqref{eq:nm1} and \ref{eq:nm2} for the definition of the number and anomalous densities, as a 
functional derivative  of the expectation value of an observable, is what distinguishes my approach from previous 
approaches in literature.  One can easily show that in the canonical basis, 

\begin{align}
\langle \Phi|\Phi(\eta_0)\rangle =
\prod_{k=1}^{2\Omega} \sqrt{1+(e^{2i\eta_0}-1)n_k } 
\end{align}
and where the canonical occupation numbers $n_k$ are double degenerate. 
For this reason there is no singularity in Eqs.~\eqref{eq:NNF} and \eqref{eq:NNF1} when
$1+(e^{2i\eta_0}-1)n_k=0$ only when both $\eta_0=\pm\pi/2$ and $n_k=1/2$.
For $\eta=0$ one obtains the corresponding unprojected densities. 
Formulas for projected densities on both the total and fragment numbers are straightforward to derive.

Notice that the qpwfs 
\begin{align}
\sum_k {\textrm u}_k(\xi){\textrm u}_k^*(\xi')+ {\textrm v}_k(\xi){\textrm v}_k^*(\xi')=\delta(\xi-\xi')
\end{align}
form a complete non-orthogonal set. This holds true for the qpwfs in the canonical basis as well. 

It is useful as well to define the projected density matrix respectively
\begin{align}
&n(\xi,\xi'|N) = \frac{1}{P(N)} {\text Re} \int_0^\pi\frac{d \eta_0}{\pi} e^{-i\eta_0N}   n(\xi,\xi'|\eta_0), \\
& N = \int d\xi \, n(\xi,\xi|N),\\
& \sum_{k=0}^{2\Omega} n_k  = \sum_{N=0}^{2\Omega} N P(N),
\end{align}
which as expected has the correct normalization.

As discussed in Ref.~\cite{Bulgac:2019a} the densities~\eqref{eq:NNF}  and \eqref{eq:NNF1}
can be used to evaluate the number projected energy of a system as follows
\begin{align}
&E(N) = \frac{1}{P(N)}{\text Re}\int_0^{\pi}\!\! \frac{d\eta_0}{\pi} e^{-i\eta_0 N }\\
&\quad \quad \times\int d\xi \, {\cal E}[n(\xi,\xi |\eta_0),... ], \\
&P(N)=  {\text Re} \int_{0}^{\pi} \frac{d\eta}{\pi} e^{-iN\eta_0}\langle |\Phi|\Phi(\eta_0)\rangle,\\
&\sum_{N=0}^{2\Omega} P(N)=1,\\
&\sum_{N=0}^{2\Omega} E(N)P(N) = \int d\xi \,{\cal E}[n(\xi,\xi |\eta_0),... ]_{\eta_0=0},
\end{align}
and unlike the prescriptions suggested in the 
past~\cite{Anguiano:2001,Robledo:2007,Dobaczewski:2007,Lacroix:2009,Bender:2009,Duguet:2009,Bally:2021},
these expressions have no singularities. This aspect was discussed  
in Ref.~\cite{Bulgac:2019a}, and it is also evident from their definitions, 
as the needed overlaps to evaluate these densities and their derivatives $\langle\Phi|\Phi(\epsilon,\eta_0)\rangle$ 
have by construction no singularities.

\section{Simultaneous projection on particle number and a fission fragment intrinsic spin}\label{sec:VIII}

One can introduce the transformation of the ${\textrm v}$-components of the qpwfs when applying a projection operator. 
The overlap matrix element (for one kind of nucleons) is in this case
$\langle \Phi|\Phi(\eta_0,\eta^\text{F},\beta)\rangle $ is given by
\begin{align}
&\langle \Phi | \Phi (\eta_0,\eta^\text{F},\beta) \rangle  = \sqrt{ \det{ \left [ \delta_{kl} + O_{kl}(\eta_0,\eta^\text{F},\beta^\text{F}) \right ]  }  },\\
&O_{kl}(\eta_0,\eta^\text{F},\beta)  = \langle {\textrm v}_k | e^{2i\eta_0}e^{i\eta^\text{F}\Theta^\text{F}}e^{iJ_x^\text{F}\beta^\text{F}} - 1 | {\textrm v}_l\rangle, \label{eq:rot1}
\end{align}
using an obvious generalization of the argumentation presented in Section~\ref{sec:II}. The practical advantage of using this type 
of angular momentum operator becomes clear when one considers simulations, where nuclei are placed in rectangular boxes. 
While the ${\textrm v}$-components of the qpwfs are localized around the center of mass of a fragment and their rotated support 
remain localized in such a localized spatial domain, the ${\textrm u}$-components 
are fully delocalized~\cite{Bulgac:1980} and their rotated support is ill defined in such simulation boxes.

The intrinsic spin of corresponding fragment is  
$\bm{J}^\text{F}=\int \!\!dxdy\,  \psi^\dagger(x)\psi(y)\langle x|\bm{j}^F|y\rangle$~\cite{Bulgac:2019a}, where 
\begin{align}
&\langle x | \bm{j}^F | y\rangle \\
&=\langle x |\Theta^\text{F}(\bm{r})[(\bm{r}-\bm{R}^\text{F})\times(\bm{p}-m\bm{v}^\text{F})+\bm{s}]\Theta^F(\bm{r})|y\rangle, \nonumber
\end{align}
and $\bm{r}$ and $\bm{p}$ are the nucleon coordinate and momentum, $\bm{s}$ its spin,
$m$  the nucleon mass, $\bm{R}^\text{F}$ and $\bm{v}^\text{F}$ are 
the center of mass and the center of mass velocity of the respective FF, and $\Theta^\text{F}(\bm{r})=1$ 
only in a finite volume centered around that FF and otherwise $\Theta^\text{F}(\bm{r})\equiv 0$.  

The probability that a FF emerges with $N^\text{F}$ particle number  and total intrinsic spin $J^\text{F}$ in the fission of 
an axially symmetric even-even nucleus, is given by, see also Refs.~\cite{Bertsch:2019,Bulgac:2019a}, 
\begin{align}
P(N,N^\text{F},J^\text{F}) &= \frac{2J+1}{2}\int_{-\pi/2}^{\pi/2}\!\!\!\frac{d\eta_0}{\pi} \int_{-\pi}^{\pi}\!\!\!\frac{d\eta^\text{F}}{2\pi}
\int_0^\pi \!\!\!d\beta^\text{F}\!\!\sin\beta^\text{F} \nonumber \\
&\times\langle  \Phi |\Phi(\eta_0,\eta^\text{F},\beta^\text{F})\rangle P_J(\cos\beta^\text{F} ), \label{eq:J}
\end{align}
where $P_J(x)$ is a Legendre polynomial.
This formula has a 
straightforward extension to projecting simultaneously the particle and the intrinsic spins of both FFs using the qpwfs overlap 
\begin{align}
\langle {\textrm v}_k|e^{2i\eta_0} e^{ i\eta^\text{F}\Theta^\text{F} }    
e^{ iJ_x^\text{L} \beta^\text{L}}   e^{iJ_x^\text{H}\beta^\text{H}}-1|{\textrm v}_l\rangle ,
\end{align}
where one can use for $\text{F}$ either $\text{L}$ or $\text{H}$. These equations 
are generalizations of those used recently in Ref.~\cite{Bulgac:2021}, where particle projection and 
double FF intrinsic spins distributions were not considered.

\section{Double number projection for an one-body observable} \label{sec:IX}

Now consider the overlap $ \langle \Phi | \Phi (\epsilon, \eta_0,\eta^\text{F}) \rangle $, for the transformation
\begin{align}
&{\textrm u}_n(\xi,\epsilon,\eta) = {\textrm u}_n(\xi),\nonumber \\
&{\textrm v}_n(\xi,\epsilon,\eta_0,\eta^\text{F}) 
 = [1+ 2\epsilon \hat{Q}^\text{F}]e^{2i\eta_0}e^{ i\eta^\text{F}\Theta^\text{F} } {\textrm v}_n(\xi), \label{eq:uvqe}
\end{align}
where
\begin{align}
\hat{Q}^\text{F}= \int d\xi d\xi' \langle \xi |\Theta^\text{F}Q\Theta^\text{F}|\xi' \rangle \psi^\dagger (\xi) \psi(\xi')
\end{align}
and evaluate $q(\eta_0,\eta^\text{F})$
\begin{align}
&q(\eta_0,\eta^\text{F}) =  \left . \frac{ d\langle \Phi | \Phi (\epsilon, \eta_0,\eta_\text{F}) \rangle}{ d\epsilon} \right |_{\epsilon = 0}  \label{eq:q}   \\
& = \langle\Phi |\Phi(\eta_0,\eta^\text{F}) \rangle e^{2i\eta_0}e^{i\eta^\text{F}} \sum_{kl} 
\langle {\textrm v}_k | \hat{Q}^\text{F} | {\textrm v}_l \rangle \, a_{lk}(\eta_0,\eta^\text{F}) , \nonumber\\
&\!\!\!\!\!\delta_{km} =\sum_l [  \delta_{kl} + \langle {\textrm v}_k | e^{2 i\eta_0}e^{ i\eta^\text{F}\Theta^\text{F}} -1 | {\textrm v}_l \rangle ]
\, a_{lm}(\eta_0,\eta^\text{F}),    \label{eq:aa} 
\end{align}
and thus one can evaluate the particle number projected value of $\hat{Q}^\text{F}$ 
\begin{align}
&Q(N,N^\text{F}) \\
&= \int_{-\pi/2}^{\pi/2} \!\!\frac{d\eta_0}{\pi} \int_{-\pi}^\pi\!\! \frac{d\eta^\text{F}}{2\pi}e^{-iN\eta_0-i\eta^\text{F}N^\text{F} }q(\eta_0,\eta^\text{F}).\nonumber
\end{align}
If the overlap $ \langle\Phi|\Phi(\eta_0,\eta^\text{F})\rangle$ vanishes then the inverse matrix $a_{lm}(\eta)$ does not exist. However,
the determinant $\det{ [\delta_{kl} + \langle {\textrm v}_k|e^{2i\eta_0}e^{i\eta^\text{F}\Theta^\text{F}} (1+2\epsilon Q^\text{F}) -1|{\textrm v}_l\rangle ]}$ 
clearly has no singularity for $\epsilon=0$, which implies that all these formulas are well defined everywhere. 
The formulas for the double projected number and anomalous densities, and the total energy can be derived following the steps outlined in 
previous sections.

\section{Correlations between intrinsic spins of the fission fragments}\label{sec:X}

A quantity of great interest if the correlation between the magnitudes and 
the relative orientations of the FF intrinsic spins~\cite{Vogt:2021,Wilson:2021,Marevic:2021,Randrup:2021}.
This correlation can be evaluated by generalizing Eq.~\eqref{eq:rot1}, using the canonical basis,  to the case of two FFs
\begin{align}
&\langle \Phi | \Phi (\eta_0,\eta^\text{L},\beta^\text{L},\beta^\text{H}) \rangle  \\
&= \sqrt{ \det{ \left [ \delta_{kl} + O_{kl}(\eta_0,\eta^\text{L}, \beta^\text{L},\beta^\text{H}) \right ]  }  },\nonumber \\
&O_{kl}(\eta_0,\eta^\text{L}, \beta^\text{\L},\beta^\text{H}) \\
 &= \langle {\textrm v}_k | e^{2i\eta_0}e^{i\eta^\text{L}\Theta^\text{L}} e^{i{\bm J}^\text{L}\cdot{\bm n}^\text{L}\beta^\text{L}} 
                                                                                        e^{i{\bm J}^\text{H}\cdot{\bm n}^\text{H}\beta^\text{H}} - 1 | {\textrm v}_l\rangle,   \nonumber
 \end{align}
 where ${\bm n}^\text{L,H}$ are two independent unit vectors. Since both $N$ and $N^\text{L}$ are fixed there is no need of a projection on $N^\text{H}$.
 One can simplify the projection operator in this matrix element
 \begin{align}
& e^{2i\eta_0}e^{i\eta^\text{L}\Theta^\text{L}} e^{i{\bm J}^\text{L}\cdot{\bm n}^\text{L}\beta^\text{L}} 
                                                                        e^{i{\bm J}^\text{H}\cdot{\bm n}^\text{H}\beta^\text{H}} - 1 \\
= &(e^{2i\eta_0}-1) \nonumber\\
+& e^{2i\eta_0}\Theta^\text{L}(e^{i\eta^\text{L}} e^{i{\bm J}^\text{L}\cdot{\bm n}^\text{L}\beta^\text{L}} -1 )\nonumber \\
 + & e^{2i\eta_0}\Theta^\text{H}(e^{i{\bm J}^\text{H}\cdot{\bm n}^\text{H}\beta^\text{H}} -1).\nonumber
\end{align}                                                                   
Even without performing FF particle projections, by ignoring the dependence of this overlap on $\eta_0, \eta^\text{L}$, one can extract valuable information
about the correlations between the relative orientations of the FF intrinsic spins, using the simpler overlap
\begin{align}
&O_{kl}(\beta^\text{L},\beta^\text{H}) 
= \langle {\textrm v}_k |e^{i{\bm J}^\text{L}\cdot{\bm n}^\text{L}\beta^\text{L}} 
                                                                   e^{i{\bm J}^\text{H}\cdot{\bm n}^\text{H}\beta^\text{H}} - 1 | \tilde{\textrm v}_l\rangle \label{eq:2ang} \\
 & \!\!\!=\langle \tilde{\textrm v}_k | \Theta^\text{L}[e^{i{\bm J}^\text{L}\cdot{\bm n}^\text{L}\beta^\text{L}} -1]| {\textrm v}_l\rangle 
+  \langle{\textrm v}_k | \Theta^\text{H}[e^{i{\bm J}^\text{H}\cdot{\bm n}^\text{H}\beta^\text{H}} -1]| {\textrm v}_l\rangle \nonumber 
  \end{align}
and using a small set of relative angles ${\bm n}^\text{L}\!\cdot{\bm n}^\text{H}=\cos\beta_\text{LH}$. 
However, no difference was observed between the two cases when $\hat{\bm n}^\text{L}\cdot\hat{\bm n}^\text{H}=\pm 1$
in the work reported in Ref.~\cite{Bulgac:2021}. 

There is no advantage in this case to use the canonical basis and one can proceed exactly as in Ref.~\cite{Bulgac:2021a}
and use the original basis ${\textrm v}_{k,l}(\xi)$ for axially symmetric FFs.

\section{The orbital angular momentum in spontaneous fission}\label{sec:XI}

The spontaneous fission of $^{252}$Cf is a particularly important and very clean case to discuss. 
Since this even-even nucleus has a zero spin in its ground state the FF intrinsic spins and angular momentum satisfy the trivial relation
\begin{align}
{\bm J}^\text{L}+{\bm J}^\text{H}+ \bm{\Lambda}=0  \label{eq:SSL}
\end{align}
and the distribution of the FFs orbital angular momentum can then be extracted. One can project on the 
sum of the two FF intrinsic spins with 
\begin{align}    
&P(\Lambda) = \frac{2\Lambda+1}{2} \int_0^\pi \!\!\!d\beta\sin\beta_0 \,P_\Lambda(\cos\beta_0  )\langle\Phi|\Phi(\beta_0)\rangle, \label{eq:SSL0}\\
&\langle \Phi | \Phi(\beta_0)\rangle = \sqrt{\det{ [\delta_{kl} +{O}_{kl}^\Lambda(\beta_0) ] }},
\end{align}
where 
\begin{align}
{ O}_{kl}^\Lambda(\beta_0)&
= \sum_{{\text F}=\text{L,H}}\langle {\textrm v}_k | \Theta^\text{F}[e^{iJ_x^\text{F}\beta_0} -1]| {\textrm v}_l\rangle.
\end{align}
According to Eq.~\eqref{eq:SSL}  in the case of $^{252}$Cf one has $e^{-i\Lambda_x\beta_0} =e^{i(J_x^\text{L}+J_x^\text{H})\beta_0}$ and 
in this case the projection on $\Lambda$ is equivalent to the projection on the sum of the FF intrinsic spins, 
if the total wave function has exactly the quantum numbers $0^+$, see discussion below too. This type of projector is in fact a projector on the combined 
FF intrinsic spins.  Notice that one can  flip the sign of $\beta_0$ without any consequence.

One can also add total and fragment particle projections for more detailed information  using the following qpwfs overlaps 
\begin{align}
{O}_{kl}^\Lambda(\eta_0,\beta_0) &= (e^{2i\eta_0}-1)\langle {\textrm v}_k | {\textrm v}_l\rangle \nonumber \\ 
 &+e^{2i\eta_0}\langle {\textrm v}_k |\Theta^\text{L} (e^{iJ_x^\text{L}\beta_0 }-1)|{\textrm v}_l\rangle \nonumber \\
&+e^{2i\eta_0} \langle {\textrm v}_k |\Theta^\text{H} (e^{iJ_x^\text{H}\beta_0 }-1)|{\textrm v}_l\rangle,\\
{ O}_{kl}^\Lambda(\eta_0,\eta^\text{L},\beta_0) &= (e^{2i\eta_0}-1)\langle {\textrm v}_k |{\textrm v}_l\rangle \nonumber \\ 
 &+e^{2i\eta_0}\langle {\textrm v}_k |\Theta^\text{L} (e^{i\eta^\text{L}+iJ_x^\text{L}\beta_0 }-1)|{\textrm v}_l\rangle \nonumber \\
&+ e^{2i\eta_0}\langle {\textrm v}_k |\Theta^\text{H} (e^{iJ_x^\text{H}\beta_0 }-1)|{\textrm v}_l\rangle.
\end{align}

In the general case Eq.~\eqref{eq:SSL} should read
\begin{align}
{\bm J}^\text{L}+{\bm J}^\text{H}+ \bm{\Lambda}={\bm S}_0, \label{eq:SSL1}
\end{align}
where ${\bm S}_0$ is the initial spin of the fissioning compound nucleus and Eq.~\eqref{eq:SSL0} will provide
the probability distribution $P(|{\bm \Lambda}-{\bm S}_0|)$ only.  

One can project simultaneously on both intrinsic FF spins and the FFs orbital angular momentum using the overlap
\begin{align}
&{ O}_{kl}(\beta^\text{L}+\beta_0,\beta^\text{H}+\beta_0) \nonumber \\
&= \langle {\textrm v}_k |                       e^{i J_x^\text{L}\beta^\text{L}} 
                                                                   e^{iJ^\text{H}_x\beta^\text{H}} e^{i(J_x^\text{L}+J_x^\text{H})\beta_0}- 1 |{\textrm v}_l\rangle \nonumber\\
 & =\sum_{{\text F}=\text{L,H}}\langle {\textrm v}_k | \Theta^\text{F}[e^{iJ_x^\text{F}(\beta^\text{F}+\beta_0)} -1]| {\textrm v}_l\rangle.\label{eq:3ang}
 \end{align}
This type of overlap depends only on two angles $\beta^\text{F}+\beta_0$, where $\text{F}=\text{L,\,H}$.
  
One might consider also an additional  projection to enforce the value of total angular momentum 
${\bm S}_0$, with the rotation operator
\begin{align} 
P_0 =  e^{i(J_x^\text{L}+J_x^\text{H}+ \Lambda_x)\gamma}
\end{align}
where $\Lambda_x$ rotates the entire system around its center of mass. The result of such a combined rotation is 
a rotation of each  FF around its own center of mass by an angle $2\gamma$ due to the action of both $\Lambda_x$ and $J_x^\text{F}$, 
as well as a displacement of each FF along the $y$-axis by an amount $D^\text{F}\gamma$ for small $\gamma$, 
where $D^\text{L} = A^\text{H}D/A$ and 
$D^\text{H} = A^\text{L}D/A$ and $D$ is the FF separation and $A=A^\text{L}+A^\text{H}$. Such a combined rotation and 
displacement of the FFs will make the corresponding overlap ${\cal O}^\Lambda_{kl}(\beta,\gamma)$ an extremely narrow function 
of $\gamma$ at $\gamma=0$.   The net results is that 
the effective integration interval over $\gamma$ becomes extremely small, which will lead to a negligible correction to Eq.~\eqref{eq:SSL0}.

\section{Numerical aspects} \label{sec:XII}

The extraction of a square root from a complex number leads to two possible roots and numerically the continuity of the overlap
$\langle \Psi|\Psi(\eta)\rangle$ as a function of $\eta$ is not ensured. However, one can use the function $\texttt{unwrap}$, a function common 
in many computer languages to generate a continuous overlap. 

An ambiguity can arise sometimes however if one or more occupation probabilities 
$n_k\equiv 1/2$, in which case the overlap has a zero, but only for $\eta_0 = \pm \pi/2$, 
and thus irrelevant, as discussed before~\cite{Bulgac:2019a}

In HFB calculations one can find that very deep levels have occupations probabilities 
very close to 1, but that does not seem to lead to any 
numerical issues however in our time-dependent simulations~\cite{PPFF:2021}, 
as all our $\beta_l<1$ and they always come in pairs.

The potential vanishing of the denominator in Eqs.~\eqref{eq:NNF} and \eqref{eq:NNF1} is compensated by the vanishing of the 
overlap $\langle \Phi|\Phi(\eta)\rangle$. In the case of double particle projection the equations are a bit more involved.

As the total and fragment average particle numbers 
$\langle \Phi|\hat{N}|\Phi\rangle =\sum_k\langle {\textrm v}_k|{\textrm v}_k\rangle $ and 
$\langle \Phi|\hat{N}^\Theta|\Phi\rangle = \sum_k \langle {\textrm v}_k|\Theta|^\text{F}|{\textrm v}_k\rangle$ 
can be rather easily be evaluated, the particle projection can be performed
for particle numbers in relatively small windows around these average values only and at most one or two dozen 
integration points in each variable should suffice as for small values of $|N-\langle \hat{N}\rangle |$ the integrand has only a few oscillations.
The evaluation of fragment particle projected
values of other observables (intrinsic spin, deformation, etc.) will proceed in a similar fashion as discussed above in this text.

The great advantage of working in the canonical basis when performing a double projection is that it requires a single diagonalization of the overlap  
$\langle {\textrm v}_k|{\textrm v}_l\rangle$ and a single evaluation of the overlap matrix $\langle \tilde{\textrm v}_k | \Theta^\text{F} | \tilde{\textrm v_l}\rangle$.
The numerical evaluation of the Eq.~\eqref{eq:NNf2} and its subsequent integration of the angles $\eta_0, \eta^\text{F}$ is relatively inexpensive.

When projecting FF intrinsic spins the overlap matrix element $\langle  \Phi |\Phi(\eta_0,\eta^\text{F},\beta^\text{F})\rangle$ 
is numerically significant in a relatively small interval around  $\beta^\text{F}=0$~\cite{Bulgac:2021} and thus only a small number of 
integration points are necessary to evaluate  Eq.~\eqref{eq:J} for example. The same situation occurs as well in the case 
of projecting on both FF intrinsic spins and also on the FFs orbital angular momentum. In particular, 
the projection on FF intrinsic spins and the FFs orbital angular moment
um using the qpwfs overlap~\eqref{eq:3ang} can be evaluated 
fast using the Gauss-Legendre quadrature formulas. Since the none of these Intrinsic spins and FFs orbital angular momentum 
are larger than 50$\hbar$ for each angle one can limit the number of quadrature points to at most $n\approx 50$. That number is even further 
reduced by the fact that any qpwfs overlap is negligible for angles $\pi/3$ (radians) and then only quadrature points in the 
interval $\beta_0+\beta^\text{F}\in[0, 0.7]$, a significant reduction of the number of quadrature points.

\section{Conclusions} 

I presented a new set of formulas for restoring broken symmetries in nuclear systems. These formulas are particularly useful when performing 
static and time-dependent nuclear energy density calculations. A new qualitative element of the present formalism is the absence of 
singularities for one-body densities, which plagued previous prescriptions, see Section~\ref{sec:VII}. Even though the simultaneous restoring of the 
broken particle numbers of the total system and of the reaction fragment symmetries require multiple projections, 
they appear feasible, see recent study~\cite{Bally:2021,Bulgac:2021a}.

\vspace{0.3cm}

{\bf Acknowledgements} \\

The funding from the Office of Science, Grant No. DE-FG02-97ER41014 and also provided in part by NNSA
cooperative Agreement DE-NA0003841 is greatly appreciated. 


\providecommand{\selectlanguage}[1]{}
\renewcommand{\selectlanguage}[1]{}

\bibliography{local_fission_a}

\end{document}